\documentclass[journal]{IEEEtran}
\usepackage[ruled,vlined]{algorithm2e}
\usepackage{amsfonts}
\usepackage{tablefootnote}
\usepackage{xcolor}
\usepackage{colortbl}
\usepackage{booktabs}
\usepackage{subfigure}

\usepackage[pdftex]{graphicx}
\definecolor{a}{rgb}{0.9,0.95,0.95}%
\definecolor{b}{rgb}{0.99,0.99,0.99}%
\begin{document}

\title{Efficient Programmable Random Variate Generation Accelerator from
Sensor Noise}

\author{James~Timothy~Meech, Phillip~Stanley-Marbell,~\IEEEmembership{Senior Member,~IEEE}%

\thanks{James Timothy Meech and Phillip Stanley-Marbell are with the Department
of Electrical Engineering, University of Cambridge, Cambridge, CB3 0FA UK e-mail: jtm45@cam.ac.uk.}%
}

\maketitle

\begin{abstract}
We introduce a method for non-uniform random number generation based on sampling a physical process in a controlled environment.
We demonstrate one proof-of-concept implementation of the method, that doubles the speed of Monte Carlo integration of a univariate Gaussian.
We show that the supply voltage and temperature of the physical process must be measured and compensated for to prevent the mean and standard deviation of the random number generator from drifting.
The method we present and our detailed empirical hardware measurements demonstrate the feasibility of programmable non-uniform random variate generation from low-power sensors and the effect of ADC quantization
on the statistical qualities of the approach.
\end{abstract}


\IEEEpeerreviewmaketitle

\section{Introduction}
\label{section:introduction}
\IEEEPARstart{C}{urrent} software-based methods of non-uniform random variate generation are slow and inefficient~\cite{thomas2009comparison,EHGRVAD}.
We present a programmable system capable of generating Gaussian random variates by extracting the noise properties of a MEMS sensor and demonstrate its principle and application.
Sampling a random physical process that has a known distribution provides a continuous random variable with a theoretically unlimited sample rate.
Table~\ref{Table:gaussian} compares several state-of-the-art methods of generating non-uniform random variates.
Gaussian random variate generation is typically an order of magnitude slower and less efficient than uniform random variate generation~\cite{thomas2009comparison}.
We propose a method with potential to be superior to all of the state-of-the-art methods in terms of sample rate and efficiency,
consisting of a physical noise source and an ADC.
In a hardware implementation, the sample rate of the ADC limits the random number generation rate.

\subsection{Somewhat Related Work: Uniform Random Variates}
\noindent \textit{Uniform} random numbers are widely used in cryptography~\cite{RRS}.
The hardware \textit{non-uniform} random number generators in Table~\ref{Table:gaussian} are fundamentally different to prior work on uniform random number
generators~\cite{hennebert2013entropy,wallace2016toward,cho2020random,lee2019true,perach2019asynchronous,srinivasan20102}.
Work on uniform random number generators is often based on some non-uniform physical entropy source but
these publications do not describe the distribution of the source.
This omission makes it impossible to compare them directly to the programmable random variate accelerator (PRVA) method that we present in Section~\ref{section:methodology}.

\begin{table}[hbt]
    \centering
    \caption{Comparison of state-of-the-art PRVA methods. \textbf{PRVA}: programmable random variate accelerator.
    \textbf{CPU}: central processing unit, \textbf{GPU}: graphics processing unit, \textbf{MPPA}: massively parallel processor array,
    \textbf{FPGA}: field programmable gate array, \textbf{MR}: memristor, \textbf{PD}: photon detection,
    \textbf{RET}: resonance energy transfer, \textbf{PDI}: photodiode, \textbf{EN}: electronic noise,
    \textbf{EXP}: exponential, \textbf{*}: this work.}
    \begin{tabular}{cccccc}
    \toprule
    \rowcolor{a} \bf{Source} & \bf{Speed} & \bf{Energy} & \bf{Dist(s)} & \bf{PRVA} & \bf{Publication}                    \\
    \midrule
    \rowcolor{b} CPU         & 890 Mb/s   & 3.17 Mb/J   & Normal       & Yes       & \cite{thomas2009comparison}, 2009   \\
    \rowcolor{a} GPU         & 12.9 Gb/s  & 108 Mb/J    & Normal       & Yes       & \cite{thomas2009comparison}, 2009   \\
    \rowcolor{b} MPPA        & 860 Mb/s   & 403 Mb/J    & Normal       & Yes       & \cite{thomas2009comparison}, 2009   \\
    \rowcolor{a} FPGA        & 12.1 Gb/s  & 645 Mb/J    & Normal       & Yes       & \cite{thomas2009comparison}, 2009   \\
    \hline
    \rowcolor{b} MR          & 6000 b/s   & 120 Gb/J    & Unknown      & No        & \cite{jiang2017novel}, 2017         \\
    \rowcolor{a} PD          & 1.77 Gb/s  & -           & Normal       & No        & \cite{marangon2017source}, 2017     \\
    \rowcolor{b} RET         & 2.89 Gb/s  & 578 Gb/J    & EXP          & Yes       & \cite{zhang2018architecting}, 2018  \\
    \rowcolor{a} PDI         & 17.4 Gb/s  & -           & Husumi       & No        & \cite{avesani2018source}, 2018      \\
    \rowcolor{b} PD          & 66.0 Mb/s  & -           & Arbitrary    & Yes       & \cite{nguyen2018programmable}, 2018 \\
    \rowcolor{a} PD          & 320 Mb/s   & -           & EXP          & No        & \cite{tomasi2018model}, 2018        \\
    \rowcolor{b} FPGA        & 6.40 Gb/s  & -           & Normal       & Yes       & \cite{hu2019gaussian}, 2019         \\
    \rowcolor{a} PD          & 8.25 Gb/s  & -           & Normal       & No        & \cite{guo2019parallel}, 2019        \\
    \rowcolor{b} EN          & 13.8 kb/s\footnotemark  & 209 kb/J$^1$  & Normal       & Yes       & [*], 2019            \\
    \bottomrule
    \end{tabular}
    \normalsize

    \label{Table:gaussian}
\end{table}
\footnotetext[1]{There was no direct effort to optimize speed or energy efficiency in this work.
 Back-of-the-envelope calculations show room for $10^{5}$ and $10^{3}$ increases in speed and energy efficiency respectively.}

\subsection{Generating Non-Uniform Random Variates Is Hard}\label{hard}
\noindent The inversion and accept-reject methods are used in software for generating samples from non-uniform random variates~\cite{NURVG}. Let $U$ and $F$ be uniform and
non-uniform random variables respectively and $F^{-1}$ the analytical closed-form solution for the inverse cumulative distribution function of $F$~\cite{NURVG}.
Algorithm~\ref{Algorithm1} shows the inversion method.
The inversion method requires that $F^{-1}$ has an analytical closed-form solution.
The Gaussian distribution has no analytical closed-form solution for $F^{-1}$ so cannot be used with the inversion method~\cite{EHGRVAD}.
The accept-reject method must be used instead. Let $U$ and $F$ be independent random variables and $u$ and $f$ the densities of $U$ and $F$ on $\mathbb{R}^d$, $d$-dimensional Euclidean space.
Let $c \geq 1$ be a constant such that the condition $f(x) \leq cu(x)$ holds for all $x$.
Algorithm~\ref{Algorithm2} shows the accept-reject method for generating samples from $F$.
The accept-reject method requires more math operations than the inversion method to transform a sample~\cite{NURVG}.
This causes it to take more clock cycles to compute and therefore, more time to transform each sample.
When using the accept-reject method, samples deviating from the desired distribution are rejected~\cite{NURVG}.

\begin{algorithm}[t]

\DontPrintSemicolon
\SetAlgoLined
\KwResult{Sample from non-uniform random variate $F$}
Generate uniform [0,1] random variate $U$\;
RETURN $F \leftarrow F^{-1}(U)$\;
\caption{Inversion method.}
\label{Algorithm1}

\end{algorithm}

\begin{algorithm}[t]
\DontPrintSemicolon
\SetAlgoLined
initialization $T=2$\;
initialization $U=1$\;
\KwResult{Sample from non-uniform random variate $F$}
\Repeat{$UT \leq 1$}{
Generate uniform [0,1] random variate $U$\;
Generate uniform [0,1] random variate $F$\;
Set $T \leftarrow c \frac{f(F)}{u(F)}$\;
}
RETURN $F$\;
\caption{Accept-reject method.}
\label{Algorithm2}
\end{algorithm}

\subsection{Uses of Non-Uniform Random Variates}
\noindent Non-uniform random variate generators are fundamental to applications employing Monte Carlo methods~\cite{REGFPUED}, such as
population balance modeling of the crystallization process~\cite{CQMMMCAESPBMDMCSDCCP},
ray tracing~\cite{SSCG} and financial computing~\cite{RAMCBFS}.
When conducting Bayesian inference we must evaluate Bayes' theorem to calculate the probability that a belief $B$ is true
given new data $D$, we denote this as $P(B|D)$.
To do this we need the probability that the belief is true regardless of our data ($P(B)$), the probability that the data is true given
the belief ($P(D|B)$) and the probability that the data is true regardless of the belief ($P(D)$).
We will refer to $P(B)$ as the \textit{prior}, $P(D|B)$ as the \textit{likelihood} and
$P(D)$ as the \textit{marginal likelihood}. $P(B|D)$, the \textit{posterior} is defined as:
\begin{equation}
P(B|D) = \frac{P(D|B)P(B)}{P(D)}.
\label{equation:Bayes}
\end{equation}

We calculate the marginal likelihood by integrating the joint density $P(B,D)$~\cite{lambert2018student}.
In practice, the analytical calculation of the marginal likelihood is impossible for all but the simplest joint distributions~\cite{lambert2018student}.
We instead sample from the joint distribution $P(B,D)$ to obtain summary statistics that we can use to describe it~\cite{lambert2018student}.
These random samples from bespoke probability distributions must be produced using a non-uniform random variate generator.
This kind of computation is performed on low power embedded systems such as drones for particle filter localization~\cite{pak2015improving}.
In the particle filter algorithm Equation~\ref{equation:Bayes} must be evaluated once every time step~\cite{doucet2001introduction}.

\subsection{Contributions}
\begin{enumerate}
  \item The idea that physical noise sources such as MEMS sensors can be used in a PRVA (Section~\ref{section:introduction}).
  \item Estimation of performance increase and error reduction achieved by using a PRVA for Monte Carlo integration (Section~\ref{section:MotivatingExample}).
  \item Investigation of the impact of temperature and supply voltage on the noise distribution obtained from a commercial MEMS sensor (Section~\ref{section:ThermalChamberExperiment}).
\end{enumerate}

\section{Motivating Example}
\label{section:MotivatingExample}
\noindent We performed Monte Carlo integration of a Gaussian with a mean of $\mu = 980.794$ and standard deviation of $\sigma = 7.178$ using samples from a Gaussian with the same mean and variance.
We ran the experiment with a Gaussian generated by the C++ random library and repeated it with samples from the PRVA collected at 3\,V and 20\,$^\circ$C.
The sensor-generated random variates were saved in a file and then presented to the C++ benchmark program in a $10^6$ element array.
We uniformly-interpolate between the points in the PRVA generated Gaussian using a [-1, 1] uniform C++ random number generator.
We assume that the PRVA can produce a sample in the same amount of time that it takes to perform a read from memory.
The current PRVA cannot do this but it is possible with fast ADCs and custom analog-out accelerometers. \par
\begin{figure}[t]
    \centering
    \includegraphics[width=0.45\textwidth]{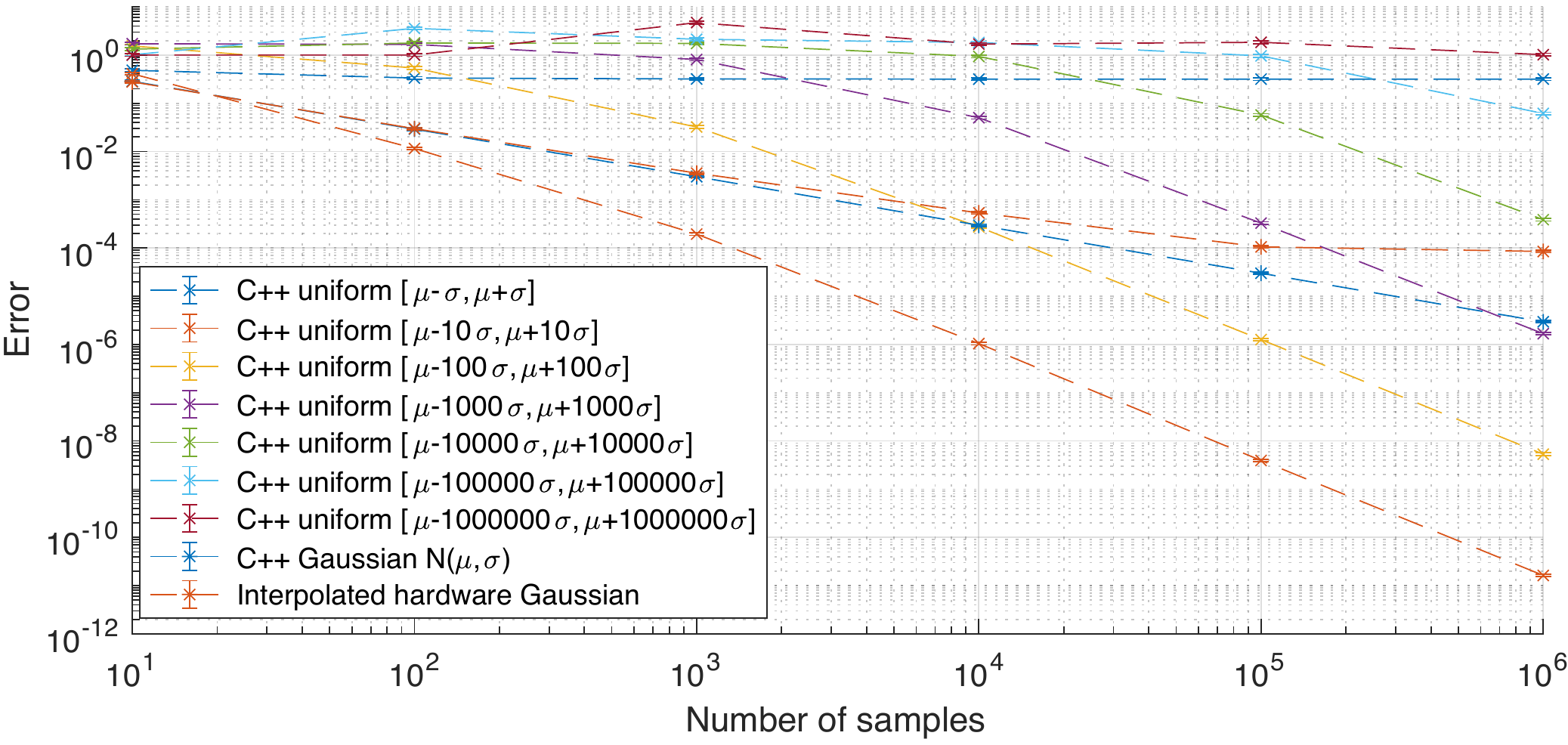}
    \caption{The error of Monte Carlo integration depends upon the range for uniform random numbers but not for Gaussian random numbers.
    The error bars were plotted using a 90\,\% confidence interval on the mean.
    The code was compiled with g++-mp-7 c++11 and run on a 2.8\,GHz Intel Core i7 CPU with Openmp to utilize all eight processor threads.}
    \label{fig:MonteError}
\end{figure}

We performed the same integration using samples from a uniform distribution with various ranges.
Let $E$ be the error of the integration, $t$ be the time taken by the integration, $N$ be the number of random
samples, and $D$ be the distribution (either uniform or Gaussian). Let samps be the array of random variates, $A$ be the area, $b$ and $h$ the
rectangle base and height, and $f$ the probability density function of the Gaussian for integration.
Algorithm~\ref{Algorithm3} shows the integration scheme that we used. We repeated each process 1000 times and calculated the average time $t$ and error $E$.
Figure~\ref{fig:MonteError} shows that the PRVA outperforms the C++ uniform random number generator for most ranges.
It is only outperformed by uniform generators with a range of $\pm 10 \sigma$ to $\pm 1000 \sigma$.
The proportion of the uniform probability density function density overlapping the Gaussian density decreases as the range of the uniform distribution is increased.
For a given function it is impossible know beforehand which range of uniform random numbers will produce a sufficiently small bound on the error of integration.
To avoid this problem we sample from a distribution that closely matches the distribution to be integrated.
The divergence between the C++ Gaussian random number generator and the PRVA is due to the lack of unique numbers in the tails of the distribution.
The PRVA performs the task up to 1.4 times faster than the C++ Gaussian random number generator with eight threads and always twice as fast with one thread.

\begin{algorithm}[h]
\DontPrintSemicolon
\SetAlgoLined
\KwResult{Error $E$ and time $t$}
Timer start\;
Generate $N$ random samples from distribution $D$\;
Sort $N$ random samples\;
\For{All pairs of samples}{
$b$ = samps[$i$] $-$ samps[$i-1$]\;
$h$ = ($f$(samps[$i$]) $+$ $f$(samps[$i-1$]))/2\;
$A$ $\mathrel{+}=$ $b \times h$\;
}
$E$ = abs($1-A$)\;
Timer stop\;
$t$ = stop $-$ start\;
RETURN $E$, $t$\;
\caption{Monte Carlo integration.}
\label{Algorithm3}
\end{algorithm}

\section{Methodology}
\label{section:methodology}

\noindent A PRVA based on physical noise sources must have negligible drift of the mean and standard deviation over time.
Drift would cause errors in calculations using the output of the PRVA.
Any environmental parameter that causes non-negligible drift must be measured and compensated for.
We sampled the z-axis of a MEMS accelerometer (the accelerometer in the Bosch BMX055) to obtain the distributions.

\subsection{Temperature-Controlled Experiments}
\label{section:ThermalChamberExperiment}
\begin{figure}[t]
    \centering
    \includegraphics[width=0.45\textwidth]{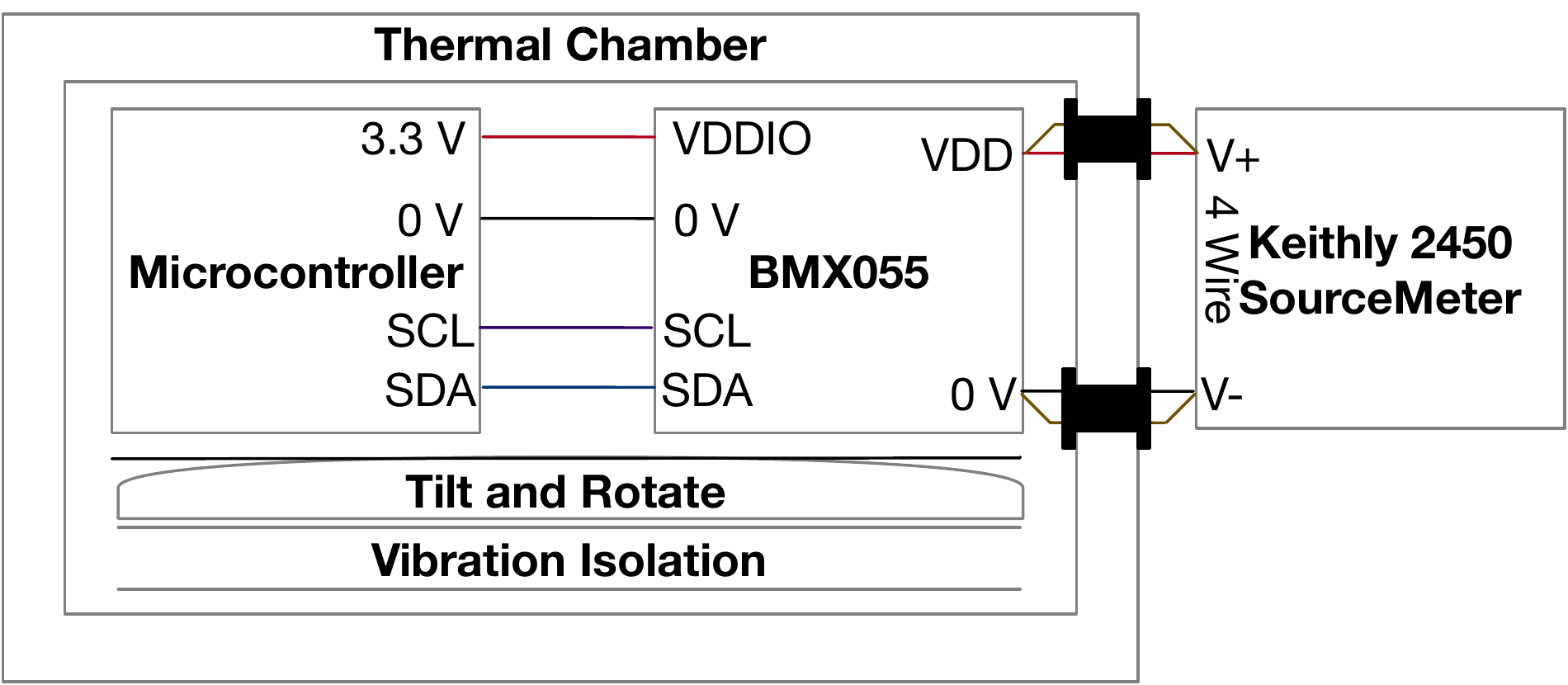}
    \caption{Experimental setup, powering the sensor and the I2C interface separately allowed more accurate measurement of the power consumed by the sensor.
             The microcontroller consumed approximately 66\,mW from its USB supply.}
    \label{Figure:CircuitDiagram}
\end{figure}

\noindent Figure~\ref{Figure:CircuitDiagram} shows the experimental setup. We placed the microcontroller, accelerometer, tilt and rotate stage and vibration isolation platform inside a Binder MK56 thermal chamber.
We connected a microcontroller to the sensor via I2C for a 1154\,Hz sample rate.
Orders of magnitude higher sample rates are possible using an off-the-shelf ADC and analog-out accelerometer.
We used a Keithly 2450 source measurement unit to power the sensor and measure the current drawn.
We set the chamber temperature to 25\,$^\circ$C and allowed 30 minutes for the temperature of the sensor to equilibrate whilst constantly sampling z-axis acceleration values from it.
We then sampled $10^5$ values from the BMX055 sensor at a 3.6\,V supply voltage.
We repeated this for all the voltages in the range of 3.6 to 1.4\,V with a 0.2\,V decrement.
We then repeated this process for the temperature from 25 down to -5\,$^\circ$C with a decrement of 5\,$^\circ$C.

\subsection{Quantization Investigation}
\noindent We investigated the effect of quantization on the Kullback–Leibler (KL) divergence between a discrete distribution and its ideal fitted curve.
We used the MATLAB \texttt{normrnd} function to generate $10^5$ values from a Gaussian distribution with the same mean and standard deviation
as the BMX055 z-axis at 2.6\,V and 10\,$^\circ$C.
We then discretized the values into various numbers of bins, fitted a Gaussian distribution to them and calculated the KL divergence between the fitted distribution and the actual distribution.

\section{Results and Discussion}
\noindent We calculated the KL divergence between two discrete distributions using the following equation~\cite{kullback1997information}. Let $P$ and $Q$ be discrete probability distributions, $x$ a given sample value and $\chi$ the sample space:
\begin{equation}
    D_{KL}(P||Q) = - \sum_{x \in \chi} P(x) \log \bigg(\frac{Q(x)}{P(x)} \bigg).
\end{equation}

\begin{figure}[]
	\centering
	\includegraphics[width=0.45\textwidth]{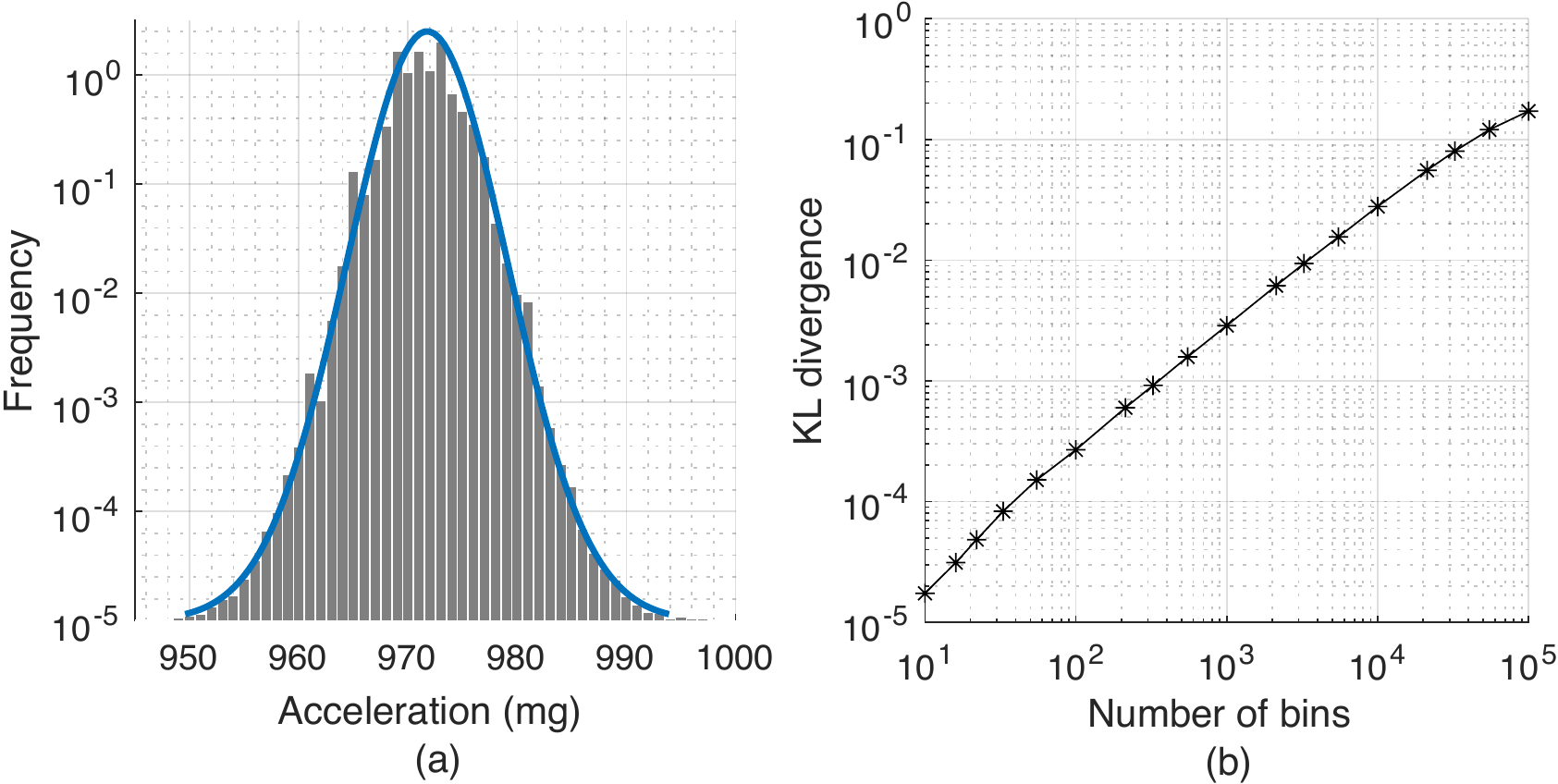}
	\caption{\textbf{(a)} Distribution of BMX055 accelerometer z-axis noise and closest fitted distribution.
  The KL divergence or difference between the data and the fitted distribution is 0.00263 demonstrating that the accelerometer noise closely matches a Gaussian distribution.
  \textbf{(b)} The effect of increasing discretization on the KL divergence whilst keeping the total number of samples constant at 100,000. Increased quantization decreases KL divergence.}
	\label{Figure:Gauss}
\end{figure}

Figure~\ref{Figure:Gauss}(a) shows the BMX055 z-axis acceleration distribution at 2.6\,V and 10\,$^\circ$C.
We found that the KL divergence between the distribution from the BMX055 z-axis and its fitted Gaussian (0.00263) was more than an order of magnitude smaller than the equivalent result for a MATLAB-generated distribution (0.0392).
This shows that the distribution of random numbers would produce accurate results in applications such as particle filters where the distribution represents the state~\cite{doucet2001introduction}.\par
We rounded the MATLAB-generated floats for comparison with the BMX055-generated integers.
The KL divergence between a $10^5$ value MATLAB-generated uniform distribution with range $[\mu-3\sigma, \mu+3\sigma]$ and its fitted Gaussian is 0.116 for reference.
We averaged the KL divergence calculations over 100 distributions to account for the random variations in the measurement.
The numbers generated by the sensor are closer to an ideal Gaussian distribution than those generated by MATLAB.
We used a custom multi-sensor embedded system to perform the initial investigation into which sensors could be used to generate non-uniform random
numbers~\cite{stanley2020warp}. We found that the BMX055 provided the highest resolution and closest fit to a Gaussian. The noise from the BME680 temperature,
pressure and humidity, HDC1000 temperature,  MMA8451 accelerometer, AMG8833 temperature and MAG3110 magnetometer sensors was also Gaussian but with a lower resolution.

\begin{figure}[]
	\centering
	\includegraphics[width=0.45\textwidth]{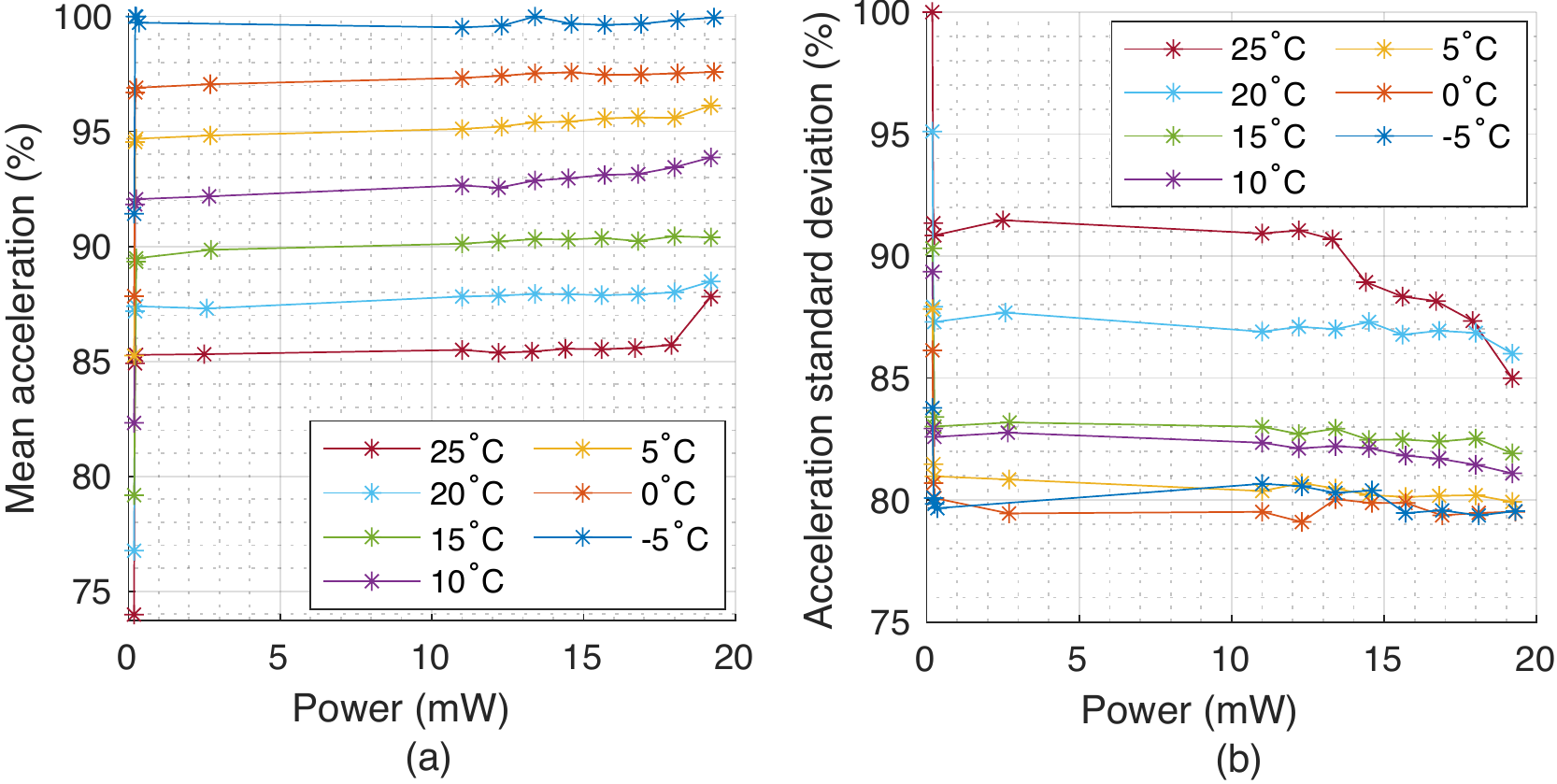}
	\caption{\textbf{(a)} The mean of BMX055 z-axis distributions decreases with increasing temperature and increases with increasing supply power. The y-axis is normalized to the maximum observed value.
  The x-axis shows the power consumed by the BMX055 sensor alone.
  \textbf{(b)} The standard deviation of BMX055 z-axis distributions decreases with increasing temperature and supply power.}
	\label{Figure:trendmean}
\end{figure}

Figures~\ref{Figure:trendmean}(a) and~\ref{Figure:trendmean}(b) show how voltage and temperature affect the mean and standard deviation of the BMX055 z-axis acceleration measurement.
Temperature has a greater effect on mean and standard deviation than voltage. Both voltage and temperature have a greater effect upon the standard deviation than the mean.
A PRVA based on this phenomenon should measure and compensate for both the temperature and the supply voltage of the sensor.
Figure~\ref{Figure:Gauss}(b) shows the effect of increasing the bin size on the divergence between a distribution of $10^5$ values and its ideal fitted distribution.
This shows that increased quantization decreases the difference between a distribution and its fitted Gaussian.

\section{Gaussian To Gaussian Transform}
\begin{figure}[t]
\begin{center}

\includegraphics[width=0.45\textwidth]{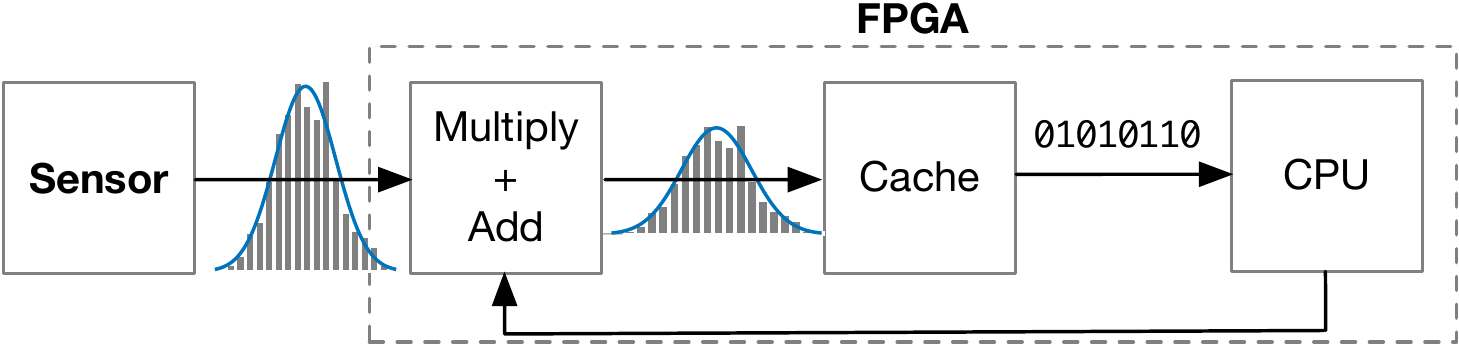}
\caption{Gaussian to Gaussian transform implementation. The CPU can request distributions with
certain parameters and they are presented to it in a cache.}
\label{Figure:GausstoGauss}
\end{center}
\end{figure}

\noindent A univariate Gaussian can be transformed to any other univariate Gaussian with one multiplication and one addition.
This is significantly less computation than the accept-reject method which requires at least 10 operations per accept-reject test:
an exponential, square, square root, subtraction, comparison and five divisions / multiplications.
The accept-reject method may need to repeat this set of operations numerous times for each random variate.
The CPU must generate two uniform random numbers as input for one accept-reject method sampling attempt.
Rapid addition and multiplication can be achieved using fast adders and multipliers implemented on an FPGA,
Figure~\ref{Figure:GausstoGauss} shows how this could be achieved.
The CPU requests a distribution by specifying parameters to the transform circuitry which proceeds to transform the input Gaussian to fit the requested output distribution.
The transform circuitry then stores the values in a small high-speed cache that the CPU can read from.
Offloading the transformation to dedicated hardware leaves the processor free to execute other instructions which will improve performance.
This architecture has not yet been implemented in an FPGA but will be the subject of future work.

\section{Conclusion}
\label{section:summary}

\noindent Sensors are a feasible entropy source for a PRVA at a higher sample rate and with greater efficiency than the state-of-the-art.
The mean and standard deviation of the noise produced by the z-axis of a commercial accelerometer depend upon the temperature of the environment and the supply voltage.
Both should be measured and compensated for in the FPGA transform circuitry.
Quantizing a Gaussian distribution decreases the KL divergence between it and a fitted Gaussian.
A PRVA can double the speed of Monte Carlo integration of a univariate Gaussian on a single thread.
The PRVA sampling rate can be increased by substituting the BMX055 for a set of analog-out accelerometers (ADXL335s) and high speed ADCs (ADS54J60s).
This method would require a customized version of the ADXL335 with the proof mass fixed in place,
amplification so the noise distribution covers at least 90\,\% of the ADC range and no low pass filtering on its output.

\ifCLASSOPTIONcaptionsoff
  \newpage
\fi

\section{Acknowledgements}

\noindent This research is supported by an Alan Turing Institute award TU/B/000096 under EPSRC grant EP/N510129/1, by EPSRC grant EP/R022534/1, by EPSRC grant EP/V004654/1, and by EPSRC grant EP/L015889/1.

\bibliography{ms}
\bibliographystyle{IEEEtran}

\end{document}